\documentstyle[psfig,11pt,a4]{article}

\newcommand{\be}{\begin{equation}}
\newcommand{\ee}{\end{equation}} 
\newcommand{\bea}{\begin{eqnarray}}
\newcommand{\eea}{\end{eqnarray}}

\begin{document}

\begin{titlepage}

\begin{flushright} 
{\tt
        SU-ITP/99-18 \\
}
 \end{flushright}

\bigskip

\begin{center}

{\bf{\LARGE 4D quantum black hole physics \\
from 2D models?
%
}}

\bigskip 
\bigskip\bigskip
 R. Balbinot$^a$ \footnote{\sc balbinot@bologna.infn.it}and A. 
 Fabbri$^b$ \footnote{Supported by an INFN fellowship. e-mail: {\sc
afabbri1@leland.stanford.edu}}

\end{center}

\bigskip%

\footnotesize
\noindent	
 a) Dipartimento di Fisica dell'Universit\`a di Bologna  
	and INFN sezione di Bologna, \\
	Via Irnerio 46, 40126 Bologna, Italy.	
 \newline                 
b) Department of Physics, Stanford University, Stanford, CA, 94305-4060, USA.
\normalsize 

\bigskip


\begin{center}
{\bf Abstract}
\end{center}
 
Minimally coupled 4D scalar fields in Schwarzschild space-time are 
considered. Dimensional reduction to 2D leads to a well known anomaly 
induced effective action, which we consider here in a local form with
the introduction of auxiliary fields. Boundary conditions are imposed on them
in order to select the appropriate quantum states (Boulware, Unruh annd 
Israel-Hartle-Hawking). The stress tensor is then calculated and 
its comparison 
with the expected 4D form turns out to be unsuccessful. 
We also critically discuss in some detail a recent controversial 
result appeared 
in the literature on the same topic.

\bigskip


\end{titlepage}

\newpage

Among the increasing variety of 2D dilaton gravity models, a particular 
attention is certainly deserved by spherically symmetric reduced 
General Relativity (GR) in 
interaction to minimally coupled massless 4D scalar fields (first considered in
\cite{MuWiZe}).  
Because of its direct link to the reald 4D world, this model has been 
regarded as a reliable device to investigate four dimensional physics 
in the s-wave sector and not just as a ``laboratory'' like other 2D models. 
The relevant effective action (or, better, a part of it) for the matter sector 
can be directly obtained by functional integration of the 2D conformal
 anomaly \cite{effac}.
This action is nonlocal, as it involves the inverse of the Delambertian 
operator. It can be made local by the introduction of two auxiliary 
fields \cite{BuRaMi}.\\
Applying a technique which has already been tested on an analogous problem in 
4D \cite{balfash}, 
in this paper we will investigate vacuum polarization around a 
Schwarzschild black hole induced by quantum minimal scalar fields using
the above mentioned effective action. 
Starting from the local form of it we will impose 
appropriate boundary conditions to the auxiliary fields in order to select
the relevant quantum states, namely Boulware (vacuum polarization around
a static star), Unruh (black hole evaporation) and Israel-Hartle-Hawking 
(thermal
equilibrium). The resulting expectation values of the stress tensor will 
be then compared to the ones obtained by canonical quantization and 
Hadamard regularization (see for instance \cite{hiansa}). 
As we shall see and as expected on the basis of previous 
results (\cite{balfab}, \cite{jorge} and references therein), 
this check dramatically fails 
and once again puts serious doubts on the possibility of inferring the actual 
4D behaviour starting from this lower dimensional effective action. 
While a similar procedure has been discussed in \cite{BuRaMi}
for the Israel-Hartle-Hawking  case, here the analysis is more general
and extended to Boulware and, in the most nontrivial case, to the Unruh state.
\\ \\ 
We shall consider 
 a $4D$ minimally coupled scalar field $f$ described
by the action 
\be
S_M^{(4)}=-\frac{1}{(4\pi)^2} \int 
d^4x \sqrt{-g^{(4)}} \partial^{\mu}f\partial_{\mu}f \ .
\label{1}
\ee
We then impose spherical symmetry both on the metric ($2+2$ splitting)
and on the field $f$, i.e.
\be
ds^2=g_{\alpha\beta}dx^{\alpha}dx^{\beta}=g_{ab}^{(2)}dx^adx^b +
e^{-2\phi(x^a)}d\Omega^2 
\label{2}
\ee
and $f=f(x^a)$ with $a,b=1,2$. By this ansatz the action (\ref{1})
becomes 
\be
S_M^{(2)}=-\frac{1}{4\pi}\int d^2x \sqrt{-g^{(2)}} 
e^{-2\phi}\partial^a f\partial_a f \ .
\label{3}
\ee
From the 2D point of view this action describes a theory for a 2D 
massless scalar field $f(x^a)$ coupled not only to the 2D geometry 
$g_{ab}^{(2)}(x^a)$ but also to a ``dilaton field'' $\phi(x^a)$ 
which is related to the radius of the transverse 2D sphere 
(see eq. (\ref{2})). 
Note that the action $S_M^{(2)}$ is invariant under conformal transformations
of the 2D metric $g_{ab}(x^a)$. This implies the vanishing of the
trace of the corresponding classical 2D stress tensor.   
At the quantum level this feature is lost and we have an anomaly 
\be
\langle T\rangle =\frac{1}{24\pi}\left[R-6(\nabla\phi)^2+6 \Box \phi
\right]\ .
\label{xxiii}
\ee
One can functionally integrate the trace anomaly (\ref{xxiii}) to obtain the 
anomaly induced effective action $S^{eff}_{an}$ for this 2D dilaton 
gravity model \cite{MuWiZe}, \cite{effac}
\be
S^{eff}_{an}=-\frac{1}{2\pi}\int d^2 x \sqrt{-g}\left[\frac{1}{48} R
\frac{1}{\Box}R
-\frac{1}{4}(\nabla\phi)^2 \frac{1}{\Box} R + \frac{1}{4}\phi R\right]\ ,
\label{xxvii}
\ee
where we recognize the Polyakov action besides other dilaton dependent 
contributions, reflecting the structure of eq. (\ref{xxiii}).
One can transform this action to a local form by introducing two auxiliary 
fields $\psi$ and $\chi$ \cite{BuRaMi}
\be
S^{eff}_{an}=-\frac{1}{96\pi}\int d^2 x \sqrt{-g}\left[ 2R(\psi -6\chi)
+(\nabla\psi)^2 -12\nabla\psi\nabla\chi-12\psi (\nabla\phi)^2+12R\phi
\right] \ .
\label{uuuu}
\ee
The auxiliary fields $\psi$ and $\chi$ satisfy the equations
\bea
\Box \psi &=& R\ , \nonumber \\
\Box \chi &=& (\nabla\phi)^2 \ , 
\label{iuppi}
\eea
which inserted in (\ref{uuuu}) recast the action in the nonlocal form 
(\ref{xxvii}). \\
The ``anomaly induced 2D stress tensor'' is defined as 
\bea
 \langle T_{ab}\rangle &=& \frac{2}{\sqrt{-g}}
\frac{\delta S^{eff}_{an}}{\delta g^{ab}}  = -\frac{1}{48\pi } \Big\{
2\nabla_{a}\nabla_{b}
\psi  -\nabla_{a} \psi\nabla_b \psi -g_{ab}\left[2R - \frac{1}{2}
 \nabla^c \psi\nabla_{c} \psi \right]\Big\} \nonumber \\
&- &  \frac{1}{4\pi}
\Big\{ -\frac{g_{ab}}{2}\left[ (\nabla\phi)^2\psi +\nabla^c\psi\nabla_c\chi
-2(\nabla\phi)^2 \right] +\nabla_a\phi\nabla_b\phi\ \psi \nonumber \\
&+&  \frac{1}{2}\left[ \nabla_{a}
\chi\nabla_{b}\psi +  \nabla_{b}\chi\nabla_a\psi\right] 
-\nabla_{a}\nabla_{b}\chi\Big\}
+\frac{1}{4\pi}
\left(g_{ab}\Box\phi -\nabla_{a}\nabla_{b}\phi\right)  \ .
\label{aiea}
\eea
Our strategy will be to solve the auxiliary field equations (\ref{iuppi})
in a Schwarzschild black hole background, insert the solutions in eqs. 
(\ref{aiea}) to find the 2D $\langle T_{ab}\rangle$ in the three states
$|B\rangle$, $|U\rangle$ and $|H\rangle$. Here again (see \cite{balfash})
the key point will be 
a careful choice of the arbitrary constants entering the homogeneous 
equations $\Box\psi=0$, $\Box\chi=0$. \\
The background is the following
\be
ds^2=g_{ab}dx^adx^b=-(1-2M/r)dt^2 + (1-2M/r)^{-1}dr^2\ , \ \phi=-\ln r \ .
\label{schz}
\ee
We see that this, by eq. (\ref{2}), corresponds to the 4D Schwarzschild 
metric. Note also that eqs. (\ref{schz}) extremise the gravitational action 
\be
S_{cl}=\frac{1}{8\pi}\int d^4 x\sqrt{-g^{(4)}}R^{(4)}=
\frac{1}{2\pi }\int d^2 x \sqrt{-g^{(2)}}e^{-2\phi}[R^{(2)} + 
2(\nabla\phi)^2 +2 e^{2\phi}] \ ,
\label{eire}
\ee  
where use of the ansatz (\ref{2}) has been made. \\
Using eqs. (\ref{schz}) the equations of motion for the auxiliary fields read
\bea
\Box \psi &=& \frac{4M}{r^3} \ , \nonumber \\
\Box \chi &=& (1-\frac{2M}{r})\frac{1}{r^2} \ , 
\label{liu}
\eea
where $\Box$ is the Delambertian for the 2D Schwarzschild metric. 
The general solution of these equations that will be relevant for our 
purposes  is
\bea
\psi &=&  C(r + 2M \ln \frac{(r-2M)}{l}) -\ln \frac{r-2M}{r} \ ,\nonumber \\
\chi &=&  bt + Dr +(2MD-\frac{1}{2})\ln \frac{(r-2M)}{l} 
-\frac{1}{2}\ln \frac{r}{l} 
\label{gesolu}
\eea
where $(l,C,D,b)$ are arbitrary constants. 
The presence of a linear term in $t$ in $\chi$ has already been explained in
 \cite{balfash}: 
it allows the possibility of having $\langle T_{rt}\rangle
\neq 0$ but still $\partial_t \langle T_{rt}\rangle =0$. 
No such term is present in $\psi$. A rapid look to eqs. (\ref{aiea})
gives the reason: $\langle T_{ab}\rangle$ depends also 
on $\psi$ and not just on
derivatives of the auxiliary fields. Therefore 
$\partial_t \langle S_{ab}\rangle =0$ requires $\partial_t\psi=0$. 
Substituting eqs. (\ref{gesolu}) into (\ref{aiea}) it is (we use
Eddington-Finkelstein coordinates $\{u,v\}$ where 
$u=t-r-2M\ln |r/2M -1|$, $v=t+r+2M\ln|r/2M -1|$)
\bea
&\ &\langle T_{uu} \rangle = -\frac{1}{48\pi}
\left[\frac{2M}{r^3}- \frac{3M^2}{r^4}-\frac{C^2}{4}\right]\nonumber \\
&-&\frac{1}{4\pi}\Big\{ \frac{1}{4r^2}(1-\frac{2M}{r})^2
\left[ Cr +2MC\ln \frac{r-2M}{l}-\ln \frac{r-2M}{r}\right] 
\nonumber \\
&+& \frac{C}{16Mr^2}\left[ 4M(D-b)r^2 -4Mr +4M^2\right]\Big\}\ ,
\nonumber \\
&\ &\langle T_{vv} \rangle = -\frac{1}{48\pi}
\left[\frac{2M}{r^3}- \frac{3M^2}{r^4}-\frac{C^2}{4}\right]\nonumber \\
&-&\frac{1}{4\pi}\Big\{ \frac{1}{4r^2}(1-\frac{2M}{r})^2
\left[ Cr +2MC\ln \frac{r-2M}{l}-\ln \frac{r-2M}{r}\right] \nonumber \\
&+& \frac{C}{16Mr^2}\left[ 4M(D+b)r^2 -4Mr +4M^2 \right]\Big\}\ ,
 \nonumber \\
&\ &\langle T_{uv} \rangle = \frac{1}{12\pi}(1-\frac{2M}{r})\frac{M}{r^3}
\ .
\label{limi}
\eea
Boundary conditions will then be imposed on $\psi$ and $\chi$ 
separately to select the relevant quantum states.
\\ \\ 
The Boulware state $|B\rangle$ coincides with Minkowski vacuum 
asymptotically, for which $\psi=0$ and $\chi=-\ln \frac{r}{l} $.
This limit can be achieved in eqs. (\ref{gesolu}) by setting 
$C=b=D=0$ ($l$ is an arbitrary parameter). We find 
\bea
\langle B|T_{uu}|B\rangle &=&  \langle B|T_{vv}|B\rangle =
\frac{1}{24\pi}\left[ -\frac{M}{r^3} + \frac{3}{2} \frac{M^2}{r^4}
\right] + \frac{1}{16\pi} (1-\frac{2M}{r})^2 \frac{1}{r^2}
\ln (1-\frac{2M}{r}), \nonumber \\
\langle B|T_{uv}|B\rangle &=& \frac{1}{12\pi} (1-\frac{2M}{r})\frac{M}{r^3}
 \ .
\label{bulla}
\eea
The Unruh vacuum $|U\rangle$ is constructed with modes that are regular
on the future event horizon; so we require both $\psi$ and $\chi$ to
be regular as $r\to 2M$, $t\to \infty$. Such a requirement sets 
$C=\frac{1}{2M}$ and $D-b=\frac{1}{4M}$. This latter gives $\chi\sim v$ 
on the future horizon. The remaining arbitrariness is eliminated by requiring 
$\langle T_{vv}\rangle =0$ as $r\to \infty$, i.e. there is no incoming
radiation on past null infinity. This further yields $D+b=\frac{C}{12}$.
Calculation of $\langle U|T_{ab}|U\rangle$ then gives
\bea
 \langle U|T_{uu}|U\rangle  &=&(1-\frac{2M}{r})^2
\Big\{ (768\pi M^2)^{-1}\left(\frac{4M}{r} +\frac{12M^2}{r^2}\right)
 \nonumber \\
&\ &-\frac{1}{16\pi r^2}\left(\ln\frac{r}{l}+\frac{r}{2M}\right)+
\frac{(1-6)}{768\pi M^2}\Big\}\ , \nonumber \\
 \langle U|T_{vv}|U\rangle &=& (1-\frac{2M}{r})^2\Big\{ (768\pi M^2)^{-1}
(\frac{4M}{r}+\frac{12M^2}{r^2}-5)  \nonumber \\
&\ & -\frac{1}{16\pi r^2}\left(\ln\frac{r}{l}+\frac{r}{2M}\right)\Big\}
+\frac{5}{768\pi M^2}  \ , 
\nonumber \\
 \langle U|T_{uv}|U\rangle &=& \langle B|T_{uv}|B\rangle \ .
\label{ulli}
\eea
Note the regularity of $\langle U|T_{ab}|U\rangle$ on the future horizon, 
i.e. $\langle T_{uu}\rangle \sim (r-2M)^2,\ \langle T_{vv}\rangle $ finite 
and $\langle T_{uv}\rangle \sim (r-2M)$. 
On future null infinity the luminosity of the hole 
\be
L=\frac{(1-6)}{768\pi M^2}
\label{olla}
\ee
is negative. This disappointing result was first obtained by 
 \cite{MuWiZe} (see also \cite{balfab}, \cite{jorge}). \\ 
The Israel-Hartle-Hawking state is an equilibrium state ($b=0$)
regular both on the future and past horizons. Regularity
on these surfaces of $\psi$ and $\chi$ is obtained by $C=\frac{1}{2M}$
and $D=\frac{1}{4M}$ yielding 
\bea
&\ &\langle H|T_{uu}|H\rangle =\langle H|T_{vv}|H\rangle =
\langle U|T_{uu}|U\rangle\ , \nonumber \\
&\ & \langle H|T_{uv}|H\rangle = \langle B|T_{uv}|B\rangle \ .
\label{atilo}
\eea 
One easily checks that $\langle H|T_{ab}|H\rangle$ is regular on both 
horizons, i.e. $\langle T_{uu}\rangle \sim (r-2M)^2$, 
$\langle T_{vv}\rangle \sim  (r-2M)^2$ and $\langle T_{uv}\rangle \sim (r-2M)$.
This state describes, at infinity, thermal equilibrium. However 
 the  energy density is negative
 ( $\frac{(1-6)}{768\pi M^2}$). This and the result for the Unruh state
are physically unacceptable. \\ \\ 
This would be all the story if the model is regarded just as one among 
other models of 2D dilaton gravity. However, the virtue of this particular 
model was its 4D origin. It was hoped therefore to obtain from the 2D analysis 
an insight into the real $\langle T_{\mu\nu}^{(4)}\rangle$ for minimally
coupled scalars in 4D Schwarzschild spacetime. 
The connection between 2D and 4D stress tensors is simply \cite{MuWiZe}
\be
\langle T_{ab}^{(4)}\rangle = \frac{\langle T_{ab}^{(2)}\rangle}{4\pi r^2}
\label{cone}
\ee
and furthemore the tangential pressure $P$ is given by 
\be
\langle P \rangle \equiv \langle T^{\ \theta}_{\theta}\rangle = 
\frac{1}{8\pi r^2\sqrt{-g^{(2)}}}
\frac{\delta S^{eff}_{an}}{\delta\phi}\ .
\label{pres}
\ee
From the action (\ref{uuuu}) we derive the pression
\be
\langle P\rangle = \frac{1}{64\pi^2 r^2}\left[ \frac{4M}{r^3}-2(1-\frac{2M}{r})
\frac{\partial_r\psi}{r}+ \frac{2\psi}{r^2}(1-\frac{4M}{r})\right] 
\label{fogepr}
\ee
and inserting the solution for the auxiliary field $\psi$ corresponding to the 
quantum states defined 
\be
 \langle B|T^{\ \theta}_{\theta}|B\rangle = \frac{1}{64\pi^2}
\left[ \frac{8M}{r^5}-\frac{2}{r^4}(1-\frac{4M}{r})\ln(1-\frac{2M}{r})\right]
\ , 
\label{laio}
\ee
\be
  \langle H|T^{\ \theta}_{\theta}|H\rangle = 
 \langle U|T^{\ \theta}_{\theta}|U\rangle = 
\frac{1}{64\pi^2}\left( \frac{8M}{r^5}-\frac{4}{r^4}+\frac{2}{r^4}
(1-\frac{4M}{r})\ln \frac{r}{l}\right) \ .
\label{prvasa}
\ee
So, given $\langle T_{ab}\rangle$ ($a,b=r,t$) 
from the previous calculations the corresponding $\langle T_{ab}^{(4)}\rangle$
is obtained just by dividing the results by $4\pi r^2$, whereas eqs.
(\ref{laio}) and
(\ref{prvasa}) give the remaining angular components (spherical symmetry 
requires 
$\langle T^{\ \theta}_{\theta}\rangle = \langle T^{\ \varphi}_{\varphi}
\rangle $).\\
Analytical expressions for $\langle T_{\mu\nu}\rangle$ 
in the states $|B\rangle$ and $|H\rangle$ are available. According to the 
analysis of \cite{hiansa}, $\langle B|T_{\mu}^{\ \nu}|B\rangle$ 
($\mu ,\nu =t,r,\theta,\phi$)
diverges (all components) like $(r-2M)^{-2}$ on the horizons (past and future),
whereas the asymptotic ($r\to\infty$) falloff is $O(r^{-6})$. 
Furthemore, it has been shown that $\langle H|T_{\mu\nu}|H\rangle$ 
is regular on the horizons while at infinity it has the characteristic 
form of thermal radiation in equilibrium at the Hawking temperature 
$T_H=(8\pi M)^{-1}$, namely
\be
\langle H|T_{\mu}^{\ \nu}|H\rangle \to \frac{\pi^2 }{30} T_H^4
diag (-1, 1/3, 1/3, 1/3) 
\label{cicciol}
\ee
as $r\to \infty$. Taking our results for the Boulware vacuum 
eqs. (\ref{bulla}), (\ref{laio}) we can find through the connection 
formulas eqs. (\ref{cone}), (\ref{pres}) the corresponding 
$\langle B|T_{\mu\nu}^{(4)}|B\rangle$. It is immediately seen that 
$\langle B|T_{\mu\nu}^{(4)}|B\rangle$ vanishes like $r^{-5}$ for 
$r\to\infty$ instead of the expected $r^{-6}$ behaviour. Furthemore
the pressure diverges as $\ln (1-2M/r)$ on the horizons and not like
$(r-2M)^{-2}$. So our 2D construction 
does not reproduce even qualitatively the 4D 
$\langle B|T_{\mu\nu}|B\rangle$.   
The situation for the Israel-Hartle-Hawking state $|H\rangle$
is much more dramatic. As discussed in \cite{BuRaMi}, we see that the
asymptotic behaviour of our $\langle H|T_{\mu\nu}^{(4)}|H\rangle$ is 
$(Mr)^{-2}$ instead of $M^{-4}$ (see eq. (\ref{cicciol}) ). 
Clearly, performing a spherical reduction and then quantizing is not
equivalent to the reverse procedure. \\
For the Unruh vacuum there are no analytical estimates of 
$\langle U|T_{\mu\nu}|U\rangle$. Since the 4D field equation
$\Box f=0$ and hence the modes are the same for both minimal and 
conformal massless scalars in the Schwarzschild spacetime, one expects, because
of the Bogoliubov transformation between {\it in} and {\it out} modes, 
an outgoing flux at infinity of the form 
\bea
\label{lmiu}
\langle U|T_{\mu}^{\ \nu}|U\rangle \to \frac{L}{4\pi r^2}  
\pmatrix{ -1 & -1 & 0 & 0
\cr 1 & 1 & 0 & 0
\cr 0 & 0 & 0 & 0
\cr 0 & 0 & 0 & 0 \cr}\ ,\ r\to \infty \ ,
\eea
with $L \propto M^{-2}$. Moreover $\langle U|T_{\mu}^{\ \nu}|U\rangle$
is requested to be regular (in a free falling frame) on the future 
event horizon. 
In view of the previous failures it is quite amazing to see that our 
$\langle U|T_{\mu\nu}^{(4)}|U\rangle $ constructed from eqs. (\ref{ulli}),
(\ref{prvasa}) does indeed behave asymptotically as eq. (\ref{lmiu}) 
and is regular on the future horizon. 
Neglecting the angular modes seems not to have drastic consequences 
in this case. However, we still have to face the problem of the negative 
luminosity (see eq. (\ref{olla}) ) predicted by the 2D theory.
It has been suggested that the addition of Weyl invariant (nonlocal) 
terms to $S^{eff}_{an}$ might improve the situation \cite{MuWiZe}, 
\cite{jorge}, but no definitive answer exists. \\ \\
At this point we should add that the authors of Ref. \cite{kummer}
do not share this negative feeling towards the effective action $S^{eff}_{an}$.
On the contrary, they claim that $S^{eff}_{an}$ does indeed lead to positive 
black hole luminosity in the Unruh state, given by $L=(768\pi M^2)^{-1}$,
remarkably the same value predicted by the Polyakov action. 
To understand this result, which is in so striking disagreement with 
our previous discussion and existing literature (\cite{MuWiZe}, \cite{balfab},
\cite{jorge}), 
we shall outline its derivation. 
The starting point is the conservation equations the 2D $\langle T_{ab}\rangle$
has to satisfy
\be
\nabla_a\langle T^{ab}\rangle + \frac{1}{\sqrt{-g}}
\frac{\delta S^{eff}_{an}}{\delta \phi}\nabla^b\phi = 
\nabla_a\langle T^{ab}\rangle + 8\pi e^{-2\phi}
 \langle P\rangle \nabla^b\phi =0\ ,
\label{conseq}
\ee
where use of eq. (\ref{pres}) has been made. One easily recognizes 
\cite{balfab} 
these to be the 4D conservation equations 
$\nabla^{\mu}\langle T_{\mu}^{(4)\nu}\rangle =0$. 
The physical meaning of eq. (\ref{conseq}) is clear: it links the expectation 
value of the pressure in a given quantum state to the expectation value 
of the 2D stress tensor $T_{ab}$ in the same quantum state. 
The basic assumption made in \cite{kummer} is that one is not legitimate to 
calculate $\langle T_{ab}\rangle$ by a straightforward differentiation of 
$S^{eff}_{an}$ with respect to the metric $g_{ab}$. This because of the 
subtleties involved in the correct definition of the asymptotics of the 
nonlocal operator (inverse Delambertian) entering $S^{eff}_{an}$
(see (\ref{xxvii}) ). Therefore $\langle T_{ab}\rangle$ has to be calculated
integrating the conservation equations (\ref{conseq}) once 
$\frac{\delta S^{eff}_{an}}{\delta\phi}$ is inserted into it. This functional
derivative, once $S^{eff}_{an}$ is expressed in conformal gauge 
$ds^2=- e^{2\rho}dx^+dx^-$, appears to be ``local'' namely
\be
\frac{\delta S^{eff}_{an}}{\delta\phi}=\frac{1}{2\pi} \left[
\partial_+\partial_-\rho + \partial_-(\rho\partial_+\phi) +
\partial_+(\rho\partial_-\phi)\right]
\label{impq}
\ee
and therefore eq. (\ref{conseq}) becomes
\be
\partial_{\mp}\langle T_{\pm\pm}\rangle 
+\partial_{\pm}\langle T_{+-} \rangle -2\partial_{\pm}\rho
\langle T_{+-}\rangle = \frac{\partial_{\pm}\phi}{2\pi}
 \left[
\partial_+\partial_-\rho + \partial_-(\rho\partial_+\phi) +
\partial_+(\rho\partial_-\phi)\right] \ .
\label{cein}
\ee
According to the authors of Ref. \cite{kummer} this approach, being 
entirely local, bypasses the problems mentioned above. 
Before proceeding we note that 
$\frac{\delta S^{eff}_{an}}{\delta\phi} $ 
in an arbitrary gauge is nonlocal (see eq. (\ref{fogepr}) ) and therefore is 
affected by the same problem of the correct definition of 
$\frac{1}{\Box}$. 
In a conformal gauge $\{ x^+,x^-\}$ the nonlocality is just hidden: 
different choices of conformal gauges yield different expressions for 
$\frac{\delta S^{eff}_{an}}{\delta\phi}$ 
(and hence of $\langle P\rangle$) which
are not related to each other by the laws of coordinate transformation (i.e. 
$\frac{\delta S^{eff}_{an}}{\delta\phi}$ 
does not transform as a scalar density, 
see \cite{balfab}). As experience with analogous problems 
in the Polyakov theory has taught us,
these different expressions represent just expectation 
values of $\langle P\rangle$ in different quantum states. 
Note finally that $\langle T_{+-}\rangle$ in eqs. (\ref{cein}) is fixed by the 
trace anomaly, it is state independent and transforms correctly as a tensor
\be
\langle T_{+-}\rangle =-\frac{1}{12\pi}\partial_+\partial_-\rho
+\frac{1}{4\pi}\left(
\partial_+\partial_-\phi -\partial_+\phi\partial_-\phi  \right)\ .
\label{liupo}
\ee
It is therefore clear that the choice of gauge in eq. (\ref{impq})
 becomes crucial: what is the gauge that correctly reproduces 
black hole evaporation? 
The mathematical procedure is then simple and requires just 
the insertion of the pressure term in eqs. (\ref{cein}), then
$\langle T_{\pm\pm}\rangle$ are calculated by straightforward integration 
once boundary conditions appropriate to the Unruh state are imposed. 
The choice of gauge made in \cite{kummer} is the Eddington-Finkelstein 
one
\be
ds^2=-(1-2M/r)dudv \ ,
\label{efidk}
\ee
which yields the following pressure
\be
\frac{e^{2\phi}}{8\pi\sqrt{-g}}\frac{\delta S^{eff}_{an}}{\delta\phi}
=\langle P\rangle = \frac{1}{64\pi^2}\left[ \frac{8M}{r^5}-\frac{2}{r^4}
(1-\frac{4M}{r})\ln (1-\frac{2M}{r}) \right]\ .
\label{gillo}
\ee
Inserting this in eq. (\ref{cein}) one can find $\langle T_{uu}\rangle$
by simple integration from $2M$ to $r$, i.e. 
\bea
&\ &\langle T_{uu}\rangle = -\frac{1}{48\pi}
\left(\frac{2M}{r^3}-\frac{3M^2}{r^4}\right)+\frac{1}{768\pi M^2} 
+\frac{1}{16\pi r^2}(1-\frac{2M}{r})^2 \ln (1-\frac{2M}{r}) \ , \nonumber \\
&=& \left[\frac{1}{768\pi M^2}\left(1+\frac{4M}{r}+\frac{12M^2}{r^2}\right)
+\frac{1}{16\pi r^2}\ln (1-\frac{2M}{r})\right] (1-\frac{2M}{r})^2
\label{oyiy}
\eea
from which in the limit $r\to\infty$ one obtains the announced result 
\be
\langle T_{uu}\rangle \to (768\pi M^2)^{-1} \ .
\label{announ}
\ee
In our opinion this result in view of its derivation is not unexpected.
The pressure $\langle P\rangle$ of eq. (\ref{gillo}) used in the conservation 
equations coincides with our eq. (\ref{laio}) , namely the Boulware pressure. 
Furthemore, given the structure of the conservation equations (\ref{cein}),
once the Boulware pressure is inserted in the r.h.s. the resulting 
$\langle T_{uu}\rangle$ can differ from our $\langle B|T_{uu}|B\rangle$ 
just by a constant, which the above integration procedure fixes to 
$(768\pi M^2)^{-1}$, yielding a vanishing $\langle T_{uu}\rangle$ 
on the horizon. 
We feel rather uneasy in considering the above expression as the $T_{uu}$
in the Unruh vacuum, as claimed in \cite{kummer}. From eq. (\ref{oyiy})
we see that $\langle T_{uu}\rangle$ in a free falling frame on the future 
event horizon diverges logarithmically. Also, the associated 
pressure has the same pathology. \footnote{The action used in \cite{kummer} 
differs by a local term $-(4\pi)^{-1} \int d^2 x\sqrt{-g}(\nabla\phi)^2$
from $S^{eff}_{an}$. This would give an additional contribution to 
$\langle T_{uu}\rangle$ in eq. (\ref{oyiy}), namely $-(8\pi r^2)^{-1}
(1-2M/r)^2$, which as can be seen vanishes (quadratically)  on the horizon
 and at infinity and therefore does not influence any of our conclusions.}
No such behaviour is expected 
in $\langle U|T_{\mu}^{\ \nu}|U\rangle$. The price paid to have a positive
luminosity, we feel, is too high.      


\end{document}